\documentclass[twocolumn,prl]{revtex4}

\usepackage{graphicx}
\usepackage{dcolumn}
\usepackage{bm}
\usepackage{multirow}
\usepackage{color}
\usepackage{amsmath}
\usepackage{gensymb}
\usepackage[colorlinks=true]{hyperref}

\definecolor{mblue}{rgb}{0,0.35,0.75}

\begin{document}
\title{Spectrally narrow exciton luminescence from monolayer MoS$_2$ exfoliated onto epitaxially grown hexagonal BN}

\author{E. Courtade$^1$}
\author{B. Han$^1$}
\author{S. Nakhaie$^2$}
\author{C. Robert$^1$}
\author{X. Marie$^1$}
\author{P. Renucci$^1$}
\author{T. Taniguchi$^3$}
\author{K. Watanabe$^3$}
\author{L. Geelhaar$^{2}$}
\author{J.M.J. Lopes$^{2}$}
\email{lopes@pdi-berlin.de}
\author{B. Urbaszek$^1$}
\email{urbaszek@insa-toulouse.fr}

\affiliation{%
$^1$Universit\'e de Toulouse, INSA-CNRS-UPS, LPCNO, 135 Av. Rangueil, 31077 Toulouse, France}
\affiliation{$^2$Paul-Drude-Institut f\"ur Festk\"orperelektronik, Leibniz-Institut im Forschungsverbund Berlin e.V., Hausvogteiplatz 5-7, 10117 Berlin, Germany}
\affiliation{$^3$National Institute for Materials Science, Tsukuba, Ibaraki 305-0044, Japan}

\begin{abstract}
The strong light-matter interaction in transition Metal dichalcogenides (TMDs) monolayers (MLs) is governed by robust excitons. Important progress has been made to control the dielectric environment surrounding the MLs, especially through hexagonal boron nitride (hBN) encapsulation, which drastically reduces the inhomogeneous contribution to the exciton linewidth. Most studies use exfoliated hBN from high quality flakes grown under high pressure. In this work, we show that hBN grown by molecular beam epitaxy (MBE) over a large surface area substrate has a similarly positive impact on the optical emission from TMD MLs. We deposit MoS$_2$ and MoSe$_2$ MLs on ultrathin hBN films (few MLs thick) grown on Ni/MgO(111) by MBE. Then we cover them with exfoliated hBN to finally obtain an encapsulated sample : exfoliated hBN/TMD ML/MBE hBN.  We observe an improved optical quality of our samples compared to TMD MLs exfoliated directly on SiO$_2$ substrates. Our results suggest that hBN grown by MBE could be used as a flat and charge free substrate for fabricating TMD-based heterostructures on a larger scale.   
\end{abstract}


\maketitle
\textit{Introduction.---} Transition metal dichalcogenides (TMDs) are van der Waals crystals with the chemical formula MX$_2$, where M is a transition metal such as Mo or W and X is a chalcogen atom such as S, Se or Te.  TMD monolayers (MLs) are direct semiconductors in the ML limit, as first shown for  MoS$_2$ in 2010  \cite{Mak:2010a,Splendiani:2010a}. Their strong light-matter interaction is governed by robust excitons \cite{Wang:2018a} with binding energies of the order of some hundreds of meV \cite{He:2014a, Ugeda:2014a, Chernikov:2014a, Ye:2014a, Qiu:2013a, Ramasubramaniam:2012a, Wang:2015b}, making them potential candidates for optoelectronic applications over a wide range of temperatures \cite{Mak:2016a}. Moreover, strong spin-orbit coupling and breaking of inversion symmetry allow to explore unique spin/valley properties in TMD MLs \cite{Xiao:2012a, Sallen:2012a, Mak:2012a, Kioseoglou:2012a, Cao:2012a, Jones:2013a, yang:2015b, Schaibley:2016a,Dey:2017a}.\\
\begin{figure}[h!]
\includegraphics[width=0.36\textwidth,keepaspectratio=true]{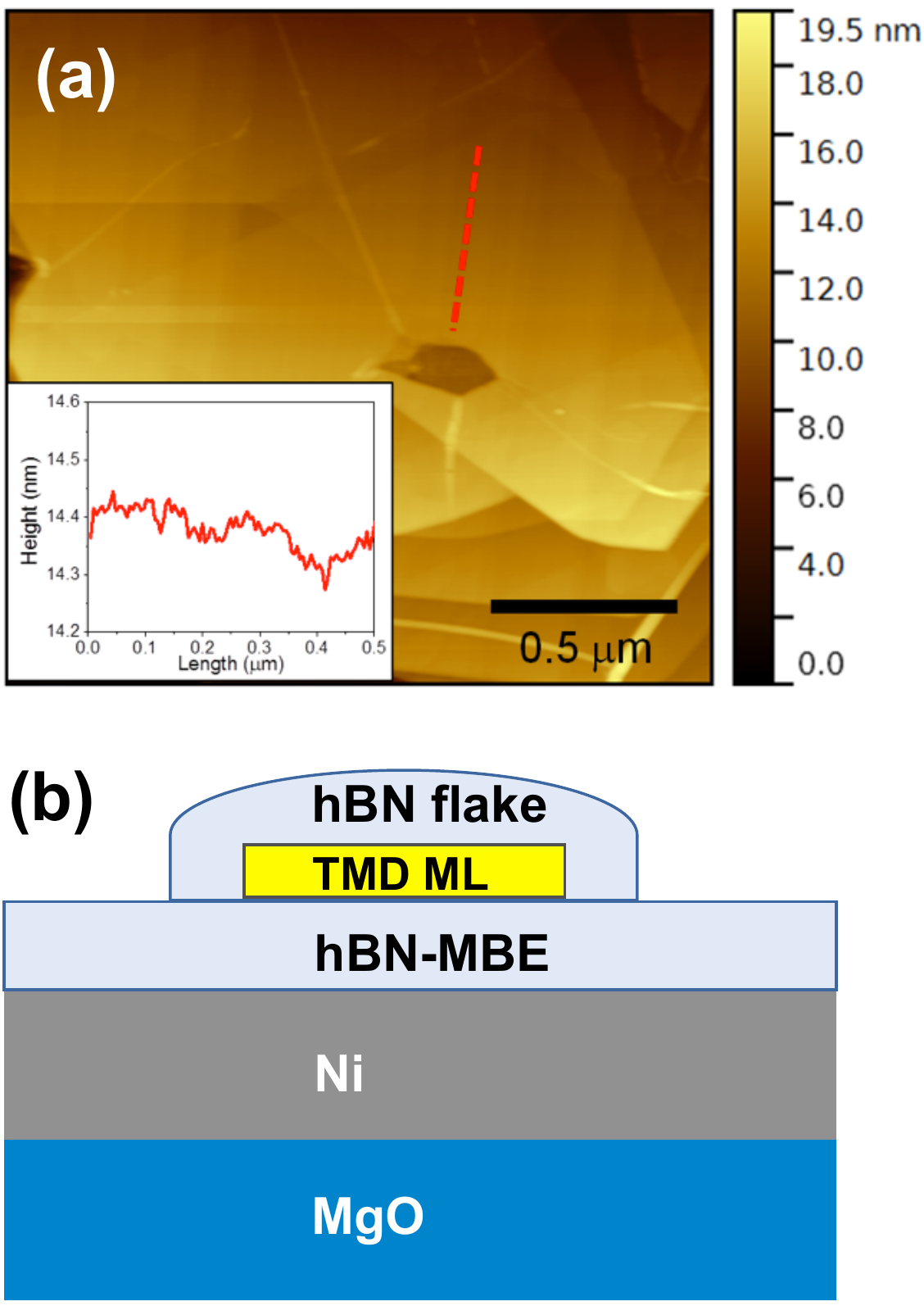}
\caption{\textbf{Investigated samples} (a) AFM image of the hBN film grown by MBE on Ni/MgO(111). The inset shows a line profile (in red) obtained across a surface region (red dashed line) that is free of wrinkles in the hBN film. (b) schematic of the investigated samples}\label{fig1} 
\end{figure}
\begin{figure*}
\includegraphics[width=0.78\textwidth,keepaspectratio=true]{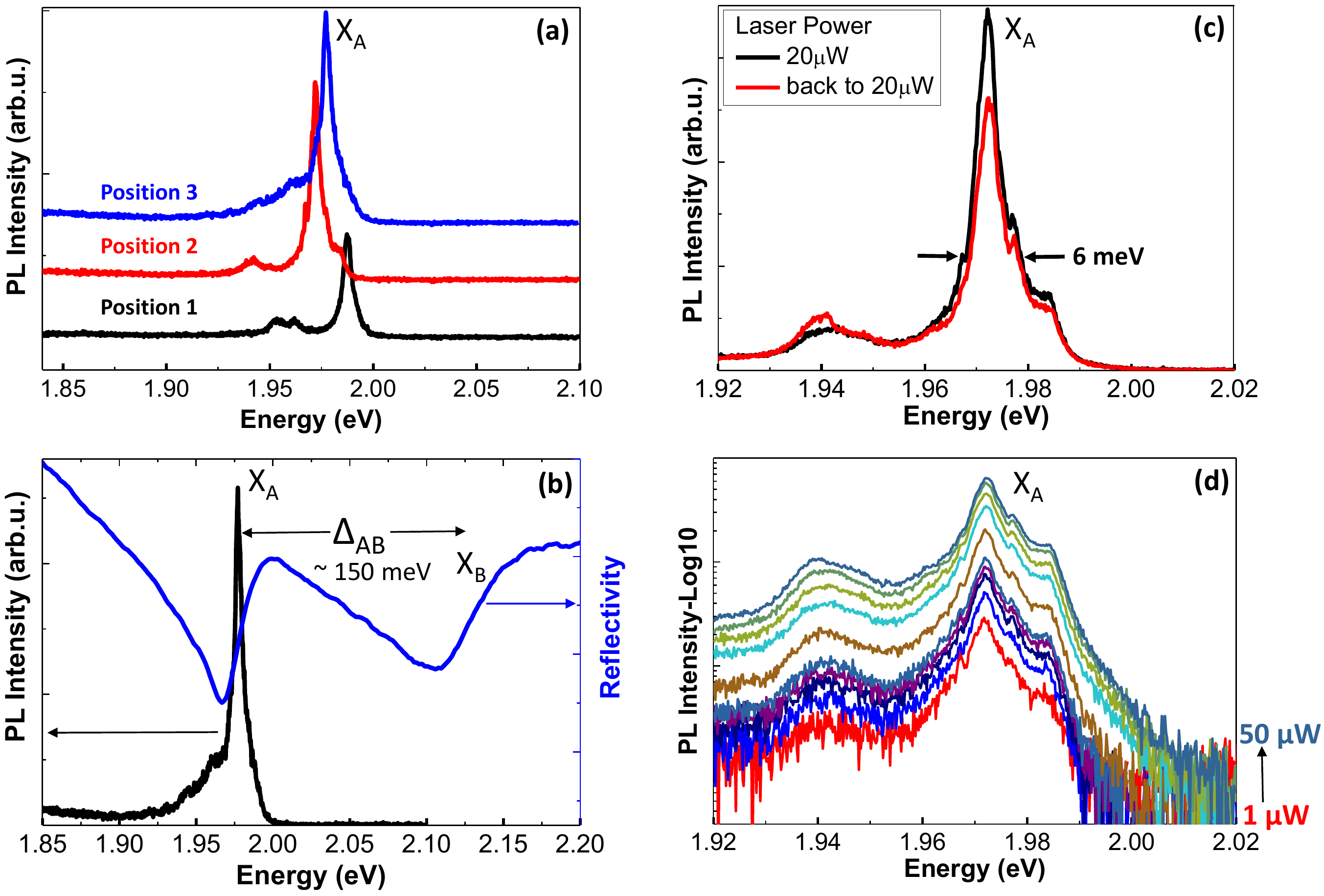}
\caption{\textbf{Optical spectroscopy results for exfoliated-hBN/MoS$_2$ ML/MBE-hBN.} (a) PL spectra with 5~$\mu$W excitation power taken at different positions to illustrate sample inhomogeneities. (b) Comparison between PL and reflectivity at the same sample position. The A exciton is visible on both spectra around 1.97~eV and the B exciton appears in reflectivity 150~meV above the A transition.  (c) PL spectra with 20~$\mu$W excitation before (black curve) and after (red curve) exposure up to 50~$\mu$W.  (d) PL spectra with increased excitation power from 1~$\mu$W to 50~$\mu$W.}\label{fig2} 
\end{figure*}
\indent For preparing field effect devices \cite{Radisavljevic:2011a}, first studies focused on TMD MLs exfoliated on SiO$_2$/Si. However, this substrate is not ideal due to surface roughness and uncontrolled charge puddles \cite{dean2010boron,britnell2012field}, responsible for very poor mobility \cite{doi:10.1021/nn401053g}. The SiO$_2$ surface is also known to degrade the optical properties of TMD MLs. The most prominent example is ML MoS$_2$, for which photoluminescence (PL) at low temperature shows neutral exciton linewidth as broad as 50~meV \cite{Korn:2011a, Zeng:2012a, Neumann:2017a, Kioseoglou:2012a, Cao:2012a, Lagarde:2014a, Sallen:2011a, stier:2015, Mitioglu:2016a}  close to a charge exciton (trion) transition and an intense defect-related emission \cite{Neumann:2017a}. 
Broad emission spectra with strong inhomogeneous contributions do not allow to take full advantage of the strong light-matter interaction and the spin-valley properties of these materials. \\
\indent Recently, TMD MLs have been encapsulated in hexagonal boron nitride (hBN) to solve this problem, similar to graphene encapsulation by hBN for transport \cite{dean2010boron,britnell2012field}. Compared to SiO$_2$, the hBN is a nearly charge free material with an atomically flat surface, the top hBN layer protects the TMD ML surface \cite{Cadiz:2017a}. These van der Waals heterostructures \cite{Geim:2013a}  are usually prepared by mechanical exfoliation. Although this is a very convenient technique for scientific research, it seems difficult to envisage device fabrication on a large scale based on exfoliation only, using flakes with dimensions in the $\mu$m to tens of $\mu$m range. That is why we investigate here substrates of 1~cm$\times$ 1~cm surface area for which molecular beam epitaxy (MBE) has been used for hBN growth instead of mechanical exfoliation. Compared to TMD MLs directly deposited on SiO$_2$/Si substrates, our samples reveal an improved optical quality. We reach neutral exciton emission linewidth as low as 6~meV for ML MoS$_2$ and 2~meV for ML MoSe$_2$, a considerable improvement in terms of FWHM compared to the same transitions for MLs deposited on SiO$_2$. We are able to distinguish optical features stemming from the A-exciton and the B-exciton state. \\ 
\indent \textit{Samples and Set-up.---} Our substrates are completely covered with an ultrathin hBN film that were grown from the constituent elements B and N using MBE. As a template for the hBN synthesis, 300 nm thick Ni films deposited on MgO(111) were employed. More details about the hBN growth procedure as well as the Ni film preparation can be found elsewhere \cite{nakhaie2015synthesis,wofford2017hybrid}. Atomic force microscopy (AFM) reveals that the hBN film offers a smooth surface morphology with a root-mean-square roughness of about 0.3 nm for a 1~$\mu$m$^{2} $ surface area (see Fig.\ref{fig1}a). The smooth nature of the surface is also illustrated in the AFM profile shown in the inset of Fig.~\ref{fig1}a. The overall surface topology shown in the AFM image is dominated by the surface features of underlying Ni film such as step clusters. Also, the existence of wrinkles in the hBN can be observed, which form during cooling due to the unequal expansion coefficients of the hBN and Ni \cite{wofford2017hybrid}. The average film thickness is 1~nm, i.e. around three MLs of hBN. Our growth technique results in crystalline hBN over the entire substrate surface. Crystalline domains are typically micrometers in diameter, as discussed in \textcite{nakhaie2015synthesis}. \\
\indent Using the all-dry viscoelastic technique described in \textcite{Gomez:2014a}, we exfoliate on the MBE grown hBN film MoS$_2$ and MoSe$_2$ monolayers. In a last step, the TMD MLs were covered with flakes exfoliated from high quality hBN grown under high pressure \cite{Taniguchi:2007a}, used in our previous studies \cite{Cadiz:2017a,Manca:2017a} to obtain encapsulated TMD MLs. A schematic representation of the prepared samples is depicted in Fig.\ref{fig1}b. The heterostructures were then investigated by performing PL and reflectivity measurements at low temperature (T=10~K) in a low vibration, closed cycle cryostat. The confocal set-up has a detection/excitation spot of about 1~$\mu$m diameter \cite{cadiz2018electrical}. To observe PL, we excited MoS$_2$ with a cw laser at 532~nm and MoSe$_2$ with a laser at cw 633~nm, whereas we used polychromatic white light for reflectivity experiments. \\
\begin{figure*}
\includegraphics[width=0.95\textwidth,keepaspectratio=true]{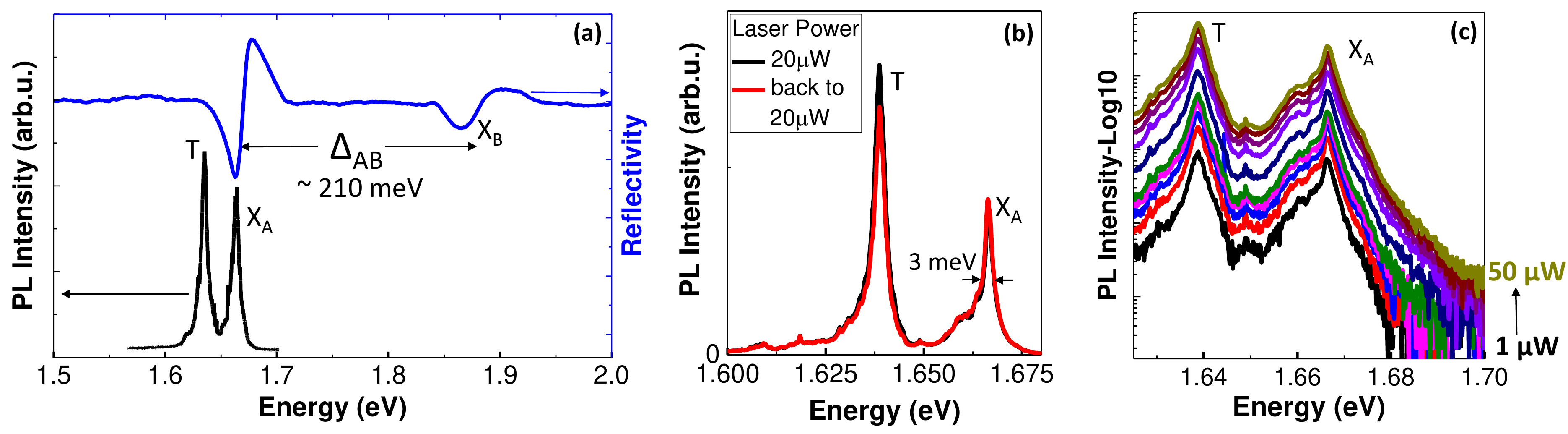}
\caption{\textbf{Optical spectroscopy results for exfoliated hBN/MoSe$_2$ ML/MBE-hBN.} (a)  Comparison between PL and reflectivity spectra. A exciton (X$_A$) is visible on both curves around 1.66~eV. In PL an additional feature usually ascribed to trion T appears at 1.635~eV. In reflectivity B exciton emerges 210~meV above the A exciton. (b) PL spectra with 20~$\mu$W excitation before (black curve) and after (red curve) exposure up to 50~$\mu$W.  (c) PL spectra with increased excitation power from 1~$\mu$W to 50~$\mu$W. }\label{fig3} 
\end{figure*}
\indent \textit{Results and Discussion.---}  First we show results for an MoS$_2$ ML covered with exfoliated hBN on top of MBE-hBN/Ni/MgO(111). By using an Attocube nano-positioner (nm step-size) we were able to choose precisely the position where we performed measurements on the encapsulated monolayer. As shown in Fig.~\ref{fig2}a, we  obtained similar, sharp PL spectra at different positions on the flake. Depending on the exact position of the detection spot on the MoS$_2$ ML flake the energy of the neutral excitonic transition (labelled $X_A$ on the spectra) can shift on average by an energy of about 10~meV, possibly due to topological imperfections of the underlying Ni layer or small wrinkles of the hBN layer grown by MBE. Our spectroscopy setup enabled us to take at the same position PL and reflectivity spectra to compare the two measurements. The overlap of the two spectra in Fig.~\ref{fig2}b shows a negligible Stokes shift (i.e. negligible neutral exciton localization), thus revealing the good optical quality of our sample under the detection spot. In addition another feature appears in reflectivity about 150~meV above the A-exciton $X_A$, associated to the B-exciton, labelled $X_B$ on the spectrum \cite{Mak:2012a}, where the A- to B-exciton separation $\Delta_{AB}$ is mainly given by the spin-orbit splitting in the valence band \cite{Kormanyos:2015a}. Transition linewidth is another characteristic which illustrates the optical quality of our samples. For MoS$_2$, we measured in PL optical linewidths down to 6~meV for the neutral A-exciton at a temperature of 10~K (see Fig.~\ref{fig2}c), which is at least 10~meV smaller compared to MLs directly deposited on SiO$_2$ \cite{Cadiz:2016a}. Moreover, the defect-related PL emission generally visible for MoS$_2$ on SiO$_2$ \cite{Sallen:2012a,Mak:2012a,Zeng:2012a,Neumann:2017a} is largely suppressed. In addition, our power-dependent measurements do not reveal irreversible changes of the shape of the PL spectra (see Fig.~\ref{fig2}d). This is again in stark contrast to what was observed previously for MoS$_2$ on SiO$_2$ where laser exposure with an average power in the $\mu$W.$\mu m^{-2} $ range induces disappearance of the neutral exciton and a shift of PL to lower energy \cite{cadiz2016ultra}. We notice a small hysteresis effect visible through the decrease of the neutral exciton intensity (see Fig.~\ref{fig2}c), possibly a consequence of weak photodoping. We performed a power-dependence cycle going from 1 to 50~$\mu$W and coming back to 1~$\mu$W then we compared two spectra taken with the same excitation power (here 20~$\mu$W) during the two parts of the cycle. In Fig.~\ref{fig2}c, we measure a reduction of the $X_A$ intensity of about 25~\%.\\
\indent To study the impact of using MBE grown hBN substrates on another important TMD material in addition to MoS$_2$, we performed measurements for a MoSe$_2$ ML \cite{Ross:2013a,dufferwiel2017valley,back2017giant,Scuri:2018a} deposited on an identical substrate of MBE-hBN/Ni/MgO(111) and covered afterwards with exfoliated hBN. In Fig.~\ref{fig3}a the overlap of PL and reflectivity shows that the neutral A-exciton $X_A$ energy is identical for the two kinds of spectroscopic experiments. This negligible Stokes shift is a first indication of the good spectral quality of the sample. In reflectivity, the B-exciton $X_B$ appears 210~meV above $X_A$, in agreement with previous measurements \cite{Li:2014a,Ross:2013a}. In PL another sharp transition commonly associated to the trion (T) \cite{Ross:2013a,2017arXiv171000988L} is visible about 29~meV below $X_A$. As illustrated in Fig.~\ref{fig3}b, the small FWHM of the transitions enables to clearly distinguish these two peaks. Depending on position, neutral exciton linewidths down to 2~meV were observed. As shown in Fig.~\ref{fig3}c PL power-dependent measurements did not induce major changes of the spectral shape, but we noticed a slight hysteresis due to photodoping (Fig.~\ref{fig3}b). The spectra reveal an increase of the PL intensity ratio $X_A$/T from 0.48 to 0.63 after laser exposure.\\
\indent In conclusion, we have shown narrow excitonic emission from MoS$_2$ and MoSe$_2$ MLs exfoliated on MBE grown hBN, suggesting that this substrate material is a suitable building block for high quality van der Waals heterostructures. 
In future experiments, the tunability of the hBN thickness on an ML level by MBE does allow in principle to see if hBN can act as a barrier material between the nickel film and the TMD to study proximity effects, similar to recent work on TMDs in contact with ferromagnetic materials \cite{zhao2017enhanced,zhong2017van}. Our MBE-grown hBN can also be used in the future as a substrate material for MBE growth of transition metal dichalcogenides \cite{ohuchi1991growth,Zhang:2014a}. \\
\indent \emph{Acknowledgements.---}  
We acknowledge stimulating discussions with Matthieu Jamet. We acknowledge funding from ANR 2D-vdW-Spin and ANR VallEx. X.M. also acknowledges the Institut Universitaire de France. Growth of hexagonal boron nitride crystals was supported by the
Elemental Strategy Initiative conducted by the MEXT, Japan and the CREST
(JPMJCR15F3), JST. 



\begin{thebibliography}{52}
\expandafter\ifx\csname natexlab\endcsname\relax\def\natexlab#1{#1}\fi
\expandafter\ifx\csname bibnamefont\endcsname\relax
  \def\bibnamefont#1{#1}\fi
\expandafter\ifx\csname bibfnamefont\endcsname\relax
  \def\bibfnamefont#1{#1}\fi
\expandafter\ifx\csname citenamefont\endcsname\relax
  \def\citenamefont#1{#1}\fi
\expandafter\ifx\csname url\endcsname\relax
  \def\url#1{\texttt{#1}}\fi
\expandafter\ifx\csname urlprefix\endcsname\relax\def\urlprefix{URL }\fi
\providecommand{\bibinfo}[2]{#2}
\providecommand{\eprint}[2][]{\url{#2}}

\bibitem[{\citenamefont{Mak et~al.}(2010)\citenamefont{Mak, Lee, Hone, Shan,
  and Heinz}}]{Mak:2010a}
\bibinfo{author}{\bibfnamefont{K.~F.} \bibnamefont{Mak}},
  \bibinfo{author}{\bibfnamefont{C.}~\bibnamefont{Lee}},
  \bibinfo{author}{\bibfnamefont{J.}~\bibnamefont{Hone}},
  \bibinfo{author}{\bibfnamefont{J.}~\bibnamefont{Shan}}, \bibnamefont{and}
  \bibinfo{author}{\bibfnamefont{T.~F.} \bibnamefont{Heinz}},
  \bibinfo{journal}{Phys. Rev. Lett.} \textbf{\bibinfo{volume}{105}},
  \bibinfo{pages}{136805} (\bibinfo{year}{2010}).

\bibitem[{\citenamefont{Splendiani et~al.}(2010)\citenamefont{Splendiani, Sun,
  Zhang, Li, Kim, Chim, Galli, and Wang}}]{Splendiani:2010a}
\bibinfo{author}{\bibfnamefont{A.}~\bibnamefont{Splendiani}},
  \bibinfo{author}{\bibfnamefont{L.}~\bibnamefont{Sun}},
  \bibinfo{author}{\bibfnamefont{Y.}~\bibnamefont{Zhang}},
  \bibinfo{author}{\bibfnamefont{T.}~\bibnamefont{Li}},
  \bibinfo{author}{\bibfnamefont{J.}~\bibnamefont{Kim}},
  \bibinfo{author}{\bibfnamefont{C.-Y.} \bibnamefont{Chim}},
  \bibinfo{author}{\bibfnamefont{G.}~\bibnamefont{Galli}}, \bibnamefont{and}
  \bibinfo{author}{\bibfnamefont{F.}~\bibnamefont{Wang}},
  \bibinfo{journal}{Nano Letters} \textbf{\bibinfo{volume}{10}},
  \bibinfo{pages}{1271} (\bibinfo{year}{2010}).

\bibitem[{\citenamefont{Wang et~al.}(2018)\citenamefont{Wang, Chernikov,
  Glazov, Heinz, Marie, Amand, and Urbaszek}}]{Wang:2018a}
\bibinfo{author}{\bibfnamefont{G.}~\bibnamefont{Wang}},
  \bibinfo{author}{\bibfnamefont{A.}~\bibnamefont{Chernikov}},
  \bibinfo{author}{\bibfnamefont{M.~M.} \bibnamefont{Glazov}},
  \bibinfo{author}{\bibfnamefont{T.~F.} \bibnamefont{Heinz}},
  \bibinfo{author}{\bibfnamefont{X.}~\bibnamefont{Marie}},
  \bibinfo{author}{\bibfnamefont{T.}~\bibnamefont{Amand}}, \bibnamefont{and}
  \bibinfo{author}{\bibfnamefont{B.}~\bibnamefont{Urbaszek}},
  \bibinfo{journal}{Rev. Mod. Phys.} \textbf{\bibinfo{volume}{90}},
  \bibinfo{pages}{021001} (\bibinfo{year}{2018}).

\bibitem[{\citenamefont{He et~al.}(2014)\citenamefont{He, Kumar, Zhao, Wang,
  Mak, Zhao, and Shan}}]{He:2014a}
\bibinfo{author}{\bibfnamefont{K.}~\bibnamefont{He}},
  \bibinfo{author}{\bibfnamefont{N.}~\bibnamefont{Kumar}},
  \bibinfo{author}{\bibfnamefont{L.}~\bibnamefont{Zhao}},
  \bibinfo{author}{\bibfnamefont{Z.}~\bibnamefont{Wang}},
  \bibinfo{author}{\bibfnamefont{K.~F.} \bibnamefont{Mak}},
  \bibinfo{author}{\bibfnamefont{H.}~\bibnamefont{Zhao}}, \bibnamefont{and}
  \bibinfo{author}{\bibfnamefont{J.}~\bibnamefont{Shan}},
  \bibinfo{journal}{Phys. Rev. Lett.} \textbf{\bibinfo{volume}{113}},
  \bibinfo{pages}{026803} (\bibinfo{year}{2014}).

\bibitem[{\citenamefont{{Ugeda} et~al.}(2014)\citenamefont{{Ugeda}, {Bradley},
  {Shi}, {da Jornada}, {Zhang}, {Qiu}, {Mo}, {Hussain}, {Shen}, {Wang}
  et~al.}}]{Ugeda:2014a}
\bibinfo{author}{\bibfnamefont{M.~M.} \bibnamefont{{Ugeda}}},
  \bibinfo{author}{\bibfnamefont{A.~J.} \bibnamefont{{Bradley}}},
  \bibinfo{author}{\bibfnamefont{S.-F.} \bibnamefont{{Shi}}},
  \bibinfo{author}{\bibfnamefont{F.~H.} \bibnamefont{{da Jornada}}},
  \bibinfo{author}{\bibfnamefont{Y.}~\bibnamefont{{Zhang}}},
  \bibinfo{author}{\bibfnamefont{D.~Y.} \bibnamefont{{Qiu}}},
  \bibinfo{author}{\bibfnamefont{S.-K.} \bibnamefont{{Mo}}},
  \bibinfo{author}{\bibfnamefont{Z.}~\bibnamefont{{Hussain}}},
  \bibinfo{author}{\bibfnamefont{Z.-X.} \bibnamefont{{Shen}}},
  \bibinfo{author}{\bibfnamefont{F.}~\bibnamefont{{Wang}}},
  \bibnamefont{et~al.}, \bibinfo{journal}{Nature Materials}
  \textbf{\bibinfo{volume}{13}}, \bibinfo{pages}{1091} (\bibinfo{year}{2014}).

\bibitem[{\citenamefont{Chernikov et~al.}(2014)\citenamefont{Chernikov,
  Berkelbach, Hill, Rigosi, Li, Aslan, Reichman, Hybertsen, and
  Heinz}}]{Chernikov:2014a}
\bibinfo{author}{\bibfnamefont{A.}~\bibnamefont{Chernikov}},
  \bibinfo{author}{\bibfnamefont{T.~C.} \bibnamefont{Berkelbach}},
  \bibinfo{author}{\bibfnamefont{H.~M.} \bibnamefont{Hill}},
  \bibinfo{author}{\bibfnamefont{A.}~\bibnamefont{Rigosi}},
  \bibinfo{author}{\bibfnamefont{Y.}~\bibnamefont{Li}},
  \bibinfo{author}{\bibfnamefont{O.~B.} \bibnamefont{Aslan}},
  \bibinfo{author}{\bibfnamefont{D.~R.} \bibnamefont{Reichman}},
  \bibinfo{author}{\bibfnamefont{M.~S.} \bibnamefont{Hybertsen}},
  \bibnamefont{and} \bibinfo{author}{\bibfnamefont{T.~F.} \bibnamefont{Heinz}},
  \bibinfo{journal}{Phys. Rev. Lett.} \textbf{\bibinfo{volume}{113}},
  \bibinfo{pages}{076802} (\bibinfo{year}{2014}).

\bibitem[{\citenamefont{{Ye} et~al.}(2014)\citenamefont{{Ye}, {Cao}, {O'Brien},
  {Zhu}, {Yin}, {Wang}, {Louie}, and {Zhang}}}]{Ye:2014a}
\bibinfo{author}{\bibfnamefont{Z.}~\bibnamefont{{Ye}}},
  \bibinfo{author}{\bibfnamefont{T.}~\bibnamefont{{Cao}}},
  \bibinfo{author}{\bibfnamefont{K.}~\bibnamefont{{O'Brien}}},
  \bibinfo{author}{\bibfnamefont{H.}~\bibnamefont{{Zhu}}},
  \bibinfo{author}{\bibfnamefont{X.}~\bibnamefont{{Yin}}},
  \bibinfo{author}{\bibfnamefont{Y.}~\bibnamefont{{Wang}}},
  \bibinfo{author}{\bibfnamefont{S.~G.} \bibnamefont{{Louie}}},
  \bibnamefont{and} \bibinfo{author}{\bibfnamefont{X.}~\bibnamefont{{Zhang}}},
  \bibinfo{journal}{Nature} \textbf{\bibinfo{volume}{513}},
  \bibinfo{pages}{214} (\bibinfo{year}{2014}).

\bibitem[{\citenamefont{Qiu et~al.}(2013)\citenamefont{Qiu, da~Jornada, and
  Louie}}]{Qiu:2013a}
\bibinfo{author}{\bibfnamefont{D.~Y.} \bibnamefont{Qiu}},
  \bibinfo{author}{\bibfnamefont{F.~H.} \bibnamefont{da~Jornada}},
  \bibnamefont{and} \bibinfo{author}{\bibfnamefont{S.~G.} \bibnamefont{Louie}},
  \bibinfo{journal}{Phys. Rev. Lett.} \textbf{\bibinfo{volume}{111}},
  \bibinfo{pages}{216805} (\bibinfo{year}{2013}).

\bibitem[{\citenamefont{Ramasubramaniam}(2012)}]{Ramasubramaniam:2012a}
\bibinfo{author}{\bibfnamefont{A.}~\bibnamefont{Ramasubramaniam}},
  \bibinfo{journal}{Phys. Rev. B} \textbf{\bibinfo{volume}{86}},
  \bibinfo{pages}{115409} (\bibinfo{year}{2012}).

\bibitem[{\citenamefont{Wang et~al.}(2015)\citenamefont{Wang, Marie, Gerber,
  Amand, Lagarde, Bouet, Vidal, Balocchi, and Urbaszek}}]{Wang:2015b}
\bibinfo{author}{\bibfnamefont{G.}~\bibnamefont{Wang}},
  \bibinfo{author}{\bibfnamefont{X.}~\bibnamefont{Marie}},
  \bibinfo{author}{\bibfnamefont{I.}~\bibnamefont{Gerber}},
  \bibinfo{author}{\bibfnamefont{T.}~\bibnamefont{Amand}},
  \bibinfo{author}{\bibfnamefont{D.}~\bibnamefont{Lagarde}},
  \bibinfo{author}{\bibfnamefont{L.}~\bibnamefont{Bouet}},
  \bibinfo{author}{\bibfnamefont{M.}~\bibnamefont{Vidal}},
  \bibinfo{author}{\bibfnamefont{A.}~\bibnamefont{Balocchi}}, \bibnamefont{and}
  \bibinfo{author}{\bibfnamefont{B.}~\bibnamefont{Urbaszek}},
  \bibinfo{journal}{Phys. Rev. Lett.} \textbf{\bibinfo{volume}{114}},
  \bibinfo{pages}{097403} (\bibinfo{year}{2015}).

\bibitem[{\citenamefont{Mak and Shan}(2016)}]{Mak:2016a}
\bibinfo{author}{\bibfnamefont{K.~F.} \bibnamefont{Mak}} \bibnamefont{and}
  \bibinfo{author}{\bibfnamefont{J.}~\bibnamefont{Shan}},
  \bibinfo{journal}{Nature Photonics} \textbf{\bibinfo{volume}{10}},
  \bibinfo{pages}{216} (\bibinfo{year}{2016}).

\bibitem[{\citenamefont{Xiao et~al.}(2012)\citenamefont{Xiao, Liu, Feng, Xu,
  and Yao}}]{Xiao:2012a}
\bibinfo{author}{\bibfnamefont{D.}~\bibnamefont{Xiao}},
  \bibinfo{author}{\bibfnamefont{G.-B.} \bibnamefont{Liu}},
  \bibinfo{author}{\bibfnamefont{W.}~\bibnamefont{Feng}},
  \bibinfo{author}{\bibfnamefont{X.}~\bibnamefont{Xu}}, \bibnamefont{and}
  \bibinfo{author}{\bibfnamefont{W.}~\bibnamefont{Yao}},
  \bibinfo{journal}{Phys. Rev. Lett.} \textbf{\bibinfo{volume}{108}},
  \bibinfo{pages}{196802} (\bibinfo{year}{2012}).

\bibitem[{\citenamefont{Sallen et~al.}(2012)\citenamefont{Sallen, Bouet, Marie,
  Wang, Zhu, Han, Lu, Tan, Amand, Liu et~al.}}]{Sallen:2012a}
\bibinfo{author}{\bibfnamefont{G.}~\bibnamefont{Sallen}},
  \bibinfo{author}{\bibfnamefont{L.}~\bibnamefont{Bouet}},
  \bibinfo{author}{\bibfnamefont{X.}~\bibnamefont{Marie}},
  \bibinfo{author}{\bibfnamefont{G.}~\bibnamefont{Wang}},
  \bibinfo{author}{\bibfnamefont{C.~R.} \bibnamefont{Zhu}},
  \bibinfo{author}{\bibfnamefont{W.~P.} \bibnamefont{Han}},
  \bibinfo{author}{\bibfnamefont{Y.}~\bibnamefont{Lu}},
  \bibinfo{author}{\bibfnamefont{P.~H.} \bibnamefont{Tan}},
  \bibinfo{author}{\bibfnamefont{T.}~\bibnamefont{Amand}},
  \bibinfo{author}{\bibfnamefont{B.~L.} \bibnamefont{Liu}},
  \bibnamefont{et~al.}, \bibinfo{journal}{Phys. Rev. B}
  \textbf{\bibinfo{volume}{86}}, \bibinfo{pages}{081301}
  (\bibinfo{year}{2012}).

\bibitem[{\citenamefont{Mak et~al.}(2012)\citenamefont{Mak, He, Shan, and
  Heinz}}]{Mak:2012a}
\bibinfo{author}{\bibfnamefont{K.~F.} \bibnamefont{Mak}},
  \bibinfo{author}{\bibfnamefont{K.}~\bibnamefont{He}},
  \bibinfo{author}{\bibfnamefont{J.}~\bibnamefont{Shan}}, \bibnamefont{and}
  \bibinfo{author}{\bibfnamefont{T.~F.} \bibnamefont{Heinz}},
  \bibinfo{journal}{Nat. Nanotechnol.} \textbf{\bibinfo{volume}{7}},
  \bibinfo{pages}{494} (\bibinfo{year}{2012}).

\bibitem[{\citenamefont{Kioseoglou et~al.}(2012)\citenamefont{Kioseoglou,
  Hanbicki, Currie, Friedman, Gunlycke, and Jonker}}]{Kioseoglou:2012a}
\bibinfo{author}{\bibfnamefont{G.}~\bibnamefont{Kioseoglou}},
  \bibinfo{author}{\bibfnamefont{A.~T.} \bibnamefont{Hanbicki}},
  \bibinfo{author}{\bibfnamefont{M.}~\bibnamefont{Currie}},
  \bibinfo{author}{\bibfnamefont{A.~L.} \bibnamefont{Friedman}},
  \bibinfo{author}{\bibfnamefont{D.}~\bibnamefont{Gunlycke}}, \bibnamefont{and}
  \bibinfo{author}{\bibfnamefont{B.~T.} \bibnamefont{Jonker}},
  \bibinfo{journal}{Applied Physics Letters} \textbf{\bibinfo{volume}{101}},
  \bibinfo{eid}{221907} (pages~\bibinfo{numpages}{4}) (\bibinfo{year}{2012}).

\bibitem[{\citenamefont{Cao et~al.}(2012)\citenamefont{Cao, Wang, Han, Ye, Zhu,
  Shi, Niu, Tan, Wang, Liu et~al.}}]{Cao:2012a}
\bibinfo{author}{\bibfnamefont{T.}~\bibnamefont{Cao}},
  \bibinfo{author}{\bibfnamefont{G.}~\bibnamefont{Wang}},
  \bibinfo{author}{\bibfnamefont{W.}~\bibnamefont{Han}},
  \bibinfo{author}{\bibfnamefont{H.}~\bibnamefont{Ye}},
  \bibinfo{author}{\bibfnamefont{C.}~\bibnamefont{Zhu}},
  \bibinfo{author}{\bibfnamefont{J.}~\bibnamefont{Shi}},
  \bibinfo{author}{\bibfnamefont{Q.}~\bibnamefont{Niu}},
  \bibinfo{author}{\bibfnamefont{P.}~\bibnamefont{Tan}},
  \bibinfo{author}{\bibfnamefont{E.}~\bibnamefont{Wang}},
  \bibinfo{author}{\bibfnamefont{B.}~\bibnamefont{Liu}}, \bibnamefont{et~al.},
  \bibinfo{journal}{Nature Communications} \textbf{\bibinfo{volume}{3}},
  \bibinfo{pages}{887} (\bibinfo{year}{2012}).

\bibitem[{\citenamefont{Jones et~al.}(2013)\citenamefont{Jones, Yu, Ghimire,
  Wu, Aivazian, Ross, Zhao, Yan, Mandrus, Xiao et~al.}}]{Jones:2013a}
\bibinfo{author}{\bibfnamefont{A.~M.} \bibnamefont{Jones}},
  \bibinfo{author}{\bibfnamefont{H.}~\bibnamefont{Yu}},
  \bibinfo{author}{\bibfnamefont{N.~J.} \bibnamefont{Ghimire}},
  \bibinfo{author}{\bibfnamefont{S.}~\bibnamefont{Wu}},
  \bibinfo{author}{\bibfnamefont{G.}~\bibnamefont{Aivazian}},
  \bibinfo{author}{\bibfnamefont{J.~S.} \bibnamefont{Ross}},
  \bibinfo{author}{\bibfnamefont{B.}~\bibnamefont{Zhao}},
  \bibinfo{author}{\bibfnamefont{J.}~\bibnamefont{Yan}},
  \bibinfo{author}{\bibfnamefont{D.~G.} \bibnamefont{Mandrus}},
  \bibinfo{author}{\bibfnamefont{D.}~\bibnamefont{Xiao}}, \bibnamefont{et~al.},
  \bibinfo{journal}{Nat. Nanotechnol.} \textbf{\bibinfo{volume}{8}},
  \bibinfo{pages}{634} (\bibinfo{year}{2013}).

\bibitem[{\citenamefont{Yang et~al.}(2015)\citenamefont{Yang, Chen, McCreary,
  Jonker, Lou, and Crooker}}]{yang:2015b}
\bibinfo{author}{\bibfnamefont{L.}~\bibnamefont{Yang}},
  \bibinfo{author}{\bibfnamefont{W.}~\bibnamefont{Chen}},
  \bibinfo{author}{\bibfnamefont{K.~M.} \bibnamefont{McCreary}},
  \bibinfo{author}{\bibfnamefont{B.~T.} \bibnamefont{Jonker}},
  \bibinfo{author}{\bibfnamefont{J.}~\bibnamefont{Lou}}, \bibnamefont{and}
  \bibinfo{author}{\bibfnamefont{S.~A.} \bibnamefont{Crooker}},
  \bibinfo{journal}{Nano letters} \textbf{\bibinfo{volume}{15}},
  \bibinfo{pages}{8250} (\bibinfo{year}{2015}).

\bibitem[{\citenamefont{Schaibley et~al.}(2016)\citenamefont{Schaibley, Yu,
  Clark, Rivera, Ross, Seyler, Yao, and Xu}}]{Schaibley:2016a}
\bibinfo{author}{\bibfnamefont{J.~R.} \bibnamefont{Schaibley}},
  \bibinfo{author}{\bibfnamefont{H.}~\bibnamefont{Yu}},
  \bibinfo{author}{\bibfnamefont{G.}~\bibnamefont{Clark}},
  \bibinfo{author}{\bibfnamefont{P.}~\bibnamefont{Rivera}},
  \bibinfo{author}{\bibfnamefont{J.~S.} \bibnamefont{Ross}},
  \bibinfo{author}{\bibfnamefont{K.~L.} \bibnamefont{Seyler}},
  \bibinfo{author}{\bibfnamefont{W.}~\bibnamefont{Yao}}, \bibnamefont{and}
  \bibinfo{author}{\bibfnamefont{X.}~\bibnamefont{Xu}},
  \bibinfo{journal}{Nature Reviews Materials} \textbf{\bibinfo{volume}{1}},
  \bibinfo{pages}{16055} (\bibinfo{year}{2016}).

\bibitem[{\citenamefont{Dey et~al.}(2017)\citenamefont{Dey, Yang, Robert, Wang,
  Urbaszek, Marie, and Crooker}}]{Dey:2017a}
\bibinfo{author}{\bibfnamefont{P.}~\bibnamefont{Dey}},
  \bibinfo{author}{\bibfnamefont{L.}~\bibnamefont{Yang}},
  \bibinfo{author}{\bibfnamefont{C.}~\bibnamefont{Robert}},
  \bibinfo{author}{\bibfnamefont{G.}~\bibnamefont{Wang}},
  \bibinfo{author}{\bibfnamefont{B.}~\bibnamefont{Urbaszek}},
  \bibinfo{author}{\bibfnamefont{X.}~\bibnamefont{Marie}}, \bibnamefont{and}
  \bibinfo{author}{\bibfnamefont{S.~A.} \bibnamefont{Crooker}},
  \bibinfo{journal}{Phys. Rev. Lett.} \textbf{\bibinfo{volume}{119}},
  \bibinfo{pages}{137401} (\bibinfo{year}{2017}).

\bibitem[{\citenamefont{Radisavljevic et~al.}(2011)\citenamefont{Radisavljevic,
  Radenovic, Brivio, Giacometti, and Kis}}]{Radisavljevic:2011a}
\bibinfo{author}{\bibfnamefont{B.}~\bibnamefont{Radisavljevic}},
  \bibinfo{author}{\bibfnamefont{A.}~\bibnamefont{Radenovic}},
  \bibinfo{author}{\bibfnamefont{J.}~\bibnamefont{Brivio}},
  \bibinfo{author}{\bibfnamefont{V.}~\bibnamefont{Giacometti}},
  \bibnamefont{and} \bibinfo{author}{\bibfnamefont{A.}~\bibnamefont{Kis}},
  \bibinfo{journal}{Nature. Nanotech.} \textbf{\bibinfo{volume}{6}},
  \bibinfo{pages}{147} (\bibinfo{year}{2011}).

\bibitem[{\citenamefont{Dean et~al.}(2010)\citenamefont{Dean, Young, Meric,
  Lee, Wang, Sorgenfrei, Watanabe, Taniguchi, Kim, Shepard
  et~al.}}]{dean2010boron}
\bibinfo{author}{\bibfnamefont{C.~R.} \bibnamefont{Dean}},
  \bibinfo{author}{\bibfnamefont{A.~F.} \bibnamefont{Young}},
  \bibinfo{author}{\bibfnamefont{I.}~\bibnamefont{Meric}},
  \bibinfo{author}{\bibfnamefont{C.}~\bibnamefont{Lee}},
  \bibinfo{author}{\bibfnamefont{L.}~\bibnamefont{Wang}},
  \bibinfo{author}{\bibfnamefont{S.}~\bibnamefont{Sorgenfrei}},
  \bibinfo{author}{\bibfnamefont{K.}~\bibnamefont{Watanabe}},
  \bibinfo{author}{\bibfnamefont{T.}~\bibnamefont{Taniguchi}},
  \bibinfo{author}{\bibfnamefont{P.}~\bibnamefont{Kim}},
  \bibinfo{author}{\bibfnamefont{K.~L.} \bibnamefont{Shepard}},
  \bibnamefont{et~al.}, \bibinfo{journal}{Nature nanotechnology}
  \textbf{\bibinfo{volume}{5}}, \bibinfo{pages}{722} (\bibinfo{year}{2010}).

\bibitem[{\citenamefont{Britnell et~al.}(2012)\citenamefont{Britnell,
  Gorbachev, Jalil, Belle, Schedin, Mishchenko, Georgiou, Katsnelson, Eaves,
  Morozov et~al.}}]{britnell2012field}
\bibinfo{author}{\bibfnamefont{L.}~\bibnamefont{Britnell}},
  \bibinfo{author}{\bibfnamefont{R.}~\bibnamefont{Gorbachev}},
  \bibinfo{author}{\bibfnamefont{R.}~\bibnamefont{Jalil}},
  \bibinfo{author}{\bibfnamefont{B.}~\bibnamefont{Belle}},
  \bibinfo{author}{\bibfnamefont{F.}~\bibnamefont{Schedin}},
  \bibinfo{author}{\bibfnamefont{A.}~\bibnamefont{Mishchenko}},
  \bibinfo{author}{\bibfnamefont{T.}~\bibnamefont{Georgiou}},
  \bibinfo{author}{\bibfnamefont{M.}~\bibnamefont{Katsnelson}},
  \bibinfo{author}{\bibfnamefont{L.}~\bibnamefont{Eaves}},
  \bibinfo{author}{\bibfnamefont{S.}~\bibnamefont{Morozov}},
  \bibnamefont{et~al.}, \bibinfo{journal}{Science}
  \textbf{\bibinfo{volume}{335}}, \bibinfo{pages}{947} (\bibinfo{year}{2012}).

\bibitem[{\citenamefont{Perera et~al.}(2013)\citenamefont{Perera, Lin, Chuang,
  Chamlagain, Wang, Tan, Cheng, Tománek, and Zhou}}]{doi:10.1021/nn401053g}
\bibinfo{author}{\bibfnamefont{M.~M.} \bibnamefont{Perera}},
  \bibinfo{author}{\bibfnamefont{M.-W.} \bibnamefont{Lin}},
  \bibinfo{author}{\bibfnamefont{H.-J.} \bibnamefont{Chuang}},
  \bibinfo{author}{\bibfnamefont{B.~P.} \bibnamefont{Chamlagain}},
  \bibinfo{author}{\bibfnamefont{C.}~\bibnamefont{Wang}},
  \bibinfo{author}{\bibfnamefont{X.}~\bibnamefont{Tan}},
  \bibinfo{author}{\bibfnamefont{M.~M.-C.} \bibnamefont{Cheng}},
  \bibinfo{author}{\bibfnamefont{D.}~\bibnamefont{Tománek}}, \bibnamefont{and}
  \bibinfo{author}{\bibfnamefont{Z.}~\bibnamefont{Zhou}}, \bibinfo{journal}{ACS
  Nano} \textbf{\bibinfo{volume}{7}}, \bibinfo{pages}{4449}
  (\bibinfo{year}{2013}).

\bibitem[{\citenamefont{Korn et~al.}(2011)\citenamefont{Korn, Heydrich, Hirmer,
  Schmutzler, and Sch\"{u}ller}}]{Korn:2011a}
\bibinfo{author}{\bibfnamefont{T.}~\bibnamefont{Korn}},
  \bibinfo{author}{\bibfnamefont{S.}~\bibnamefont{Heydrich}},
  \bibinfo{author}{\bibfnamefont{M.}~\bibnamefont{Hirmer}},
  \bibinfo{author}{\bibfnamefont{J.}~\bibnamefont{Schmutzler}},
  \bibnamefont{and}
  \bibinfo{author}{\bibfnamefont{C.}~\bibnamefont{Sch\"{u}ller}},
  \bibinfo{journal}{Applied Physics Letters} \textbf{\bibinfo{volume}{99}},
  \bibinfo{eid}{102109} (\bibinfo{year}{2011}).

\bibitem[{\citenamefont{Zeng et~al.}(2012)\citenamefont{Zeng, Dai, Yao, Xiao,
  and Cui}}]{Zeng:2012a}
\bibinfo{author}{\bibfnamefont{H.}~\bibnamefont{Zeng}},
  \bibinfo{author}{\bibfnamefont{J.}~\bibnamefont{Dai}},
  \bibinfo{author}{\bibfnamefont{W.}~\bibnamefont{Yao}},
  \bibinfo{author}{\bibfnamefont{D.}~\bibnamefont{Xiao}}, \bibnamefont{and}
  \bibinfo{author}{\bibfnamefont{X.}~\bibnamefont{Cui}}, \bibinfo{journal}{Nat.
  Nanotechnol.} \textbf{\bibinfo{volume}{7}}, \bibinfo{pages}{490}
  (\bibinfo{year}{2012}).

\bibitem[{\citenamefont{Neumann et~al.}(2017)\citenamefont{Neumann, Lindlau,
  Colombier, Nutz, Najmaei, Lou, Mohite, Yamaguchi, and
  H{\"o}gele}}]{Neumann:2017a}
\bibinfo{author}{\bibfnamefont{A.}~\bibnamefont{Neumann}},
  \bibinfo{author}{\bibfnamefont{J.}~\bibnamefont{Lindlau}},
  \bibinfo{author}{\bibfnamefont{L.}~\bibnamefont{Colombier}},
  \bibinfo{author}{\bibfnamefont{M.}~\bibnamefont{Nutz}},
  \bibinfo{author}{\bibfnamefont{S.}~\bibnamefont{Najmaei}},
  \bibinfo{author}{\bibfnamefont{J.}~\bibnamefont{Lou}},
  \bibinfo{author}{\bibfnamefont{A.~D.} \bibnamefont{Mohite}},
  \bibinfo{author}{\bibfnamefont{H.}~\bibnamefont{Yamaguchi}},
  \bibnamefont{and}
  \bibinfo{author}{\bibfnamefont{A.}~\bibnamefont{H{\"o}gele}},
  \bibinfo{journal}{Nature nanotechnology} \textbf{\bibinfo{volume}{12}},
  \bibinfo{pages}{329} (\bibinfo{year}{2017}).

\bibitem[{\citenamefont{Lagarde et~al.}(2014)\citenamefont{Lagarde, Bouet,
  Marie, Zhu, Liu, Amand, Tan, and Urbaszek}}]{Lagarde:2014a}
\bibinfo{author}{\bibfnamefont{D.}~\bibnamefont{Lagarde}},
  \bibinfo{author}{\bibfnamefont{L.}~\bibnamefont{Bouet}},
  \bibinfo{author}{\bibfnamefont{X.}~\bibnamefont{Marie}},
  \bibinfo{author}{\bibfnamefont{C.~R.} \bibnamefont{Zhu}},
  \bibinfo{author}{\bibfnamefont{B.~L.} \bibnamefont{Liu}},
  \bibinfo{author}{\bibfnamefont{T.}~\bibnamefont{Amand}},
  \bibinfo{author}{\bibfnamefont{P.~H.} \bibnamefont{Tan}}, \bibnamefont{and}
  \bibinfo{author}{\bibfnamefont{B.}~\bibnamefont{Urbaszek}},
  \bibinfo{journal}{Phys. Rev. Lett.} \textbf{\bibinfo{volume}{112}},
  \bibinfo{pages}{047401} (\bibinfo{year}{2014}).

\bibitem[{\citenamefont{Sallen et~al.}(2011)\citenamefont{Sallen, Urbaszek,
  Glazov, Ivchenko, Kuroda, Mano, Kunz, Abbarchi, Sakoda, Lagarde
  et~al.}}]{Sallen:2011a}
\bibinfo{author}{\bibfnamefont{G.}~\bibnamefont{Sallen}},
  \bibinfo{author}{\bibfnamefont{B.}~\bibnamefont{Urbaszek}},
  \bibinfo{author}{\bibfnamefont{M.~M.} \bibnamefont{Glazov}},
  \bibinfo{author}{\bibfnamefont{E.~L.} \bibnamefont{Ivchenko}},
  \bibinfo{author}{\bibfnamefont{T.}~\bibnamefont{Kuroda}},
  \bibinfo{author}{\bibfnamefont{T.}~\bibnamefont{Mano}},
  \bibinfo{author}{\bibfnamefont{S.}~\bibnamefont{Kunz}},
  \bibinfo{author}{\bibfnamefont{M.}~\bibnamefont{Abbarchi}},
  \bibinfo{author}{\bibfnamefont{K.}~\bibnamefont{Sakoda}},
  \bibinfo{author}{\bibfnamefont{D.}~\bibnamefont{Lagarde}},
  \bibnamefont{et~al.}, \bibinfo{journal}{Phys. Rev. Lett.}
  \textbf{\bibinfo{volume}{107}}, \bibinfo{pages}{166604}
  (\bibinfo{year}{2011}).

\bibitem[{\citenamefont{Stier et~al.}(2015)\citenamefont{Stier, McCreary,
  Jonker, Kono, and Crooker}}]{stier:2015}
\bibinfo{author}{\bibfnamefont{A.~V.} \bibnamefont{Stier}},
  \bibinfo{author}{\bibfnamefont{K.~M.} \bibnamefont{McCreary}},
  \bibinfo{author}{\bibfnamefont{B.~T.} \bibnamefont{Jonker}},
  \bibinfo{author}{\bibfnamefont{J.}~\bibnamefont{Kono}}, \bibnamefont{and}
  \bibinfo{author}{\bibfnamefont{S.~A.} \bibnamefont{Crooker}},
  \bibinfo{journal}{arXiv preprint arXiv:1510.07022}  (\bibinfo{year}{2015}).

\bibitem[{\citenamefont{Mitioglu et~al.}(2016)\citenamefont{Mitioglu,
  Galkowski, Surrente, Klopotowski, Dumcenco, Kis, Maude, and
  Plochocka}}]{Mitioglu:2016a}
\bibinfo{author}{\bibfnamefont{A.~A.} \bibnamefont{Mitioglu}},
  \bibinfo{author}{\bibfnamefont{K.}~\bibnamefont{Galkowski}},
  \bibinfo{author}{\bibfnamefont{A.}~\bibnamefont{Surrente}},
  \bibinfo{author}{\bibfnamefont{L.}~\bibnamefont{Klopotowski}},
  \bibinfo{author}{\bibfnamefont{D.}~\bibnamefont{Dumcenco}},
  \bibinfo{author}{\bibfnamefont{A.}~\bibnamefont{Kis}},
  \bibinfo{author}{\bibfnamefont{D.~K.} \bibnamefont{Maude}}, \bibnamefont{and}
  \bibinfo{author}{\bibfnamefont{P.}~\bibnamefont{Plochocka}},
  \bibinfo{journal}{Physical Review B} \textbf{\bibinfo{volume}{93}},
  \bibinfo{pages}{165412} (\bibinfo{year}{2016}).

\bibitem[{\citenamefont{Cadiz et~al.}(2017)\citenamefont{Cadiz, Courtade,
  Robert, Wang, Shen, Cai, Taniguchi, Watanabe, Carrere, Lagarde
  et~al.}}]{Cadiz:2017a}
\bibinfo{author}{\bibfnamefont{F.}~\bibnamefont{Cadiz}},
  \bibinfo{author}{\bibfnamefont{E.}~\bibnamefont{Courtade}},
  \bibinfo{author}{\bibfnamefont{C.}~\bibnamefont{Robert}},
  \bibinfo{author}{\bibfnamefont{G.}~\bibnamefont{Wang}},
  \bibinfo{author}{\bibfnamefont{Y.}~\bibnamefont{Shen}},
  \bibinfo{author}{\bibfnamefont{H.}~\bibnamefont{Cai}},
  \bibinfo{author}{\bibfnamefont{T.}~\bibnamefont{Taniguchi}},
  \bibinfo{author}{\bibfnamefont{K.}~\bibnamefont{Watanabe}},
  \bibinfo{author}{\bibfnamefont{H.}~\bibnamefont{Carrere}},
  \bibinfo{author}{\bibfnamefont{D.}~\bibnamefont{Lagarde}},
  \bibnamefont{et~al.}, \bibinfo{journal}{Phys. Rev. X}
  \textbf{\bibinfo{volume}{7}}, \bibinfo{pages}{021026} (\bibinfo{year}{2017}).

\bibitem[{\citenamefont{Geim and Grigorieva}(2013)}]{Geim:2013a}
\bibinfo{author}{\bibfnamefont{A.~K.} \bibnamefont{Geim}} \bibnamefont{and}
  \bibinfo{author}{\bibfnamefont{I.~V.} \bibnamefont{Grigorieva}},
  \bibinfo{journal}{Nature} \textbf{\bibinfo{volume}{499}},
  \bibinfo{pages}{419} (\bibinfo{year}{2013}).

\bibitem[{\citenamefont{Nakhaie et~al.}(2015)\citenamefont{Nakhaie, Wofford,
  Schumann, Jahn, Ramsteiner, Hanke, Lopes, and
  Riechert}}]{nakhaie2015synthesis}
\bibinfo{author}{\bibfnamefont{S.}~\bibnamefont{Nakhaie}},
  \bibinfo{author}{\bibfnamefont{J.}~\bibnamefont{Wofford}},
  \bibinfo{author}{\bibfnamefont{T.}~\bibnamefont{Schumann}},
  \bibinfo{author}{\bibfnamefont{U.}~\bibnamefont{Jahn}},
  \bibinfo{author}{\bibfnamefont{M.}~\bibnamefont{Ramsteiner}},
  \bibinfo{author}{\bibfnamefont{M.}~\bibnamefont{Hanke}},
  \bibinfo{author}{\bibfnamefont{J.}~\bibnamefont{Lopes}}, \bibnamefont{and}
  \bibinfo{author}{\bibfnamefont{H.}~\bibnamefont{Riechert}},
  \bibinfo{journal}{Applied Physics Letters} \textbf{\bibinfo{volume}{106}},
  \bibinfo{pages}{213108} (\bibinfo{year}{2015}).

\bibitem[{\citenamefont{Wofford et~al.}(2017)\citenamefont{Wofford, Nakhaie,
  Krause, Liu, Ramsteiner, Hanke, Riechert, and Lopes}}]{wofford2017hybrid}
\bibinfo{author}{\bibfnamefont{J.~M.} \bibnamefont{Wofford}},
  \bibinfo{author}{\bibfnamefont{S.}~\bibnamefont{Nakhaie}},
  \bibinfo{author}{\bibfnamefont{T.}~\bibnamefont{Krause}},
  \bibinfo{author}{\bibfnamefont{X.}~\bibnamefont{Liu}},
  \bibinfo{author}{\bibfnamefont{M.}~\bibnamefont{Ramsteiner}},
  \bibinfo{author}{\bibfnamefont{M.}~\bibnamefont{Hanke}},
  \bibinfo{author}{\bibfnamefont{H.}~\bibnamefont{Riechert}}, \bibnamefont{and}
  \bibinfo{author}{\bibfnamefont{J.~M.~J.} \bibnamefont{Lopes}},
  \bibinfo{journal}{Scientific reports} \textbf{\bibinfo{volume}{7}},
  \bibinfo{pages}{43644} (\bibinfo{year}{2017}).

\bibitem[{\citenamefont{Castellanos-Gomez
  et~al.}(2014)\citenamefont{Castellanos-Gomez, Buscema, Molenaar, Singh,
  Janssen, van~der Zant, and Steele}}]{Gomez:2014a}
\bibinfo{author}{\bibfnamefont{A.}~\bibnamefont{Castellanos-Gomez}},
  \bibinfo{author}{\bibfnamefont{M.}~\bibnamefont{Buscema}},
  \bibinfo{author}{\bibfnamefont{R.}~\bibnamefont{Molenaar}},
  \bibinfo{author}{\bibfnamefont{V.}~\bibnamefont{Singh}},
  \bibinfo{author}{\bibfnamefont{L.}~\bibnamefont{Janssen}},
  \bibinfo{author}{\bibfnamefont{H.~S.~J.} \bibnamefont{van~der Zant}},
  \bibnamefont{and} \bibinfo{author}{\bibfnamefont{G.~A.}
  \bibnamefont{Steele}}, \bibinfo{journal}{2D Materials}
  \textbf{\bibinfo{volume}{1}}, \bibinfo{pages}{011002} (\bibinfo{year}{2014}).

\bibitem[{\citenamefont{Taniguchi and Watanabe}(2007)}]{Taniguchi:2007a}
\bibinfo{author}{\bibfnamefont{T.}~\bibnamefont{Taniguchi}} \bibnamefont{and}
  \bibinfo{author}{\bibfnamefont{K.}~\bibnamefont{Watanabe}},
  \bibinfo{journal}{Journal of Crystal Growth} \textbf{\bibinfo{volume}{303}},
  \bibinfo{pages}{525 } (\bibinfo{year}{2007}).

\bibitem[{\citenamefont{Manca et~al.}(2017)\citenamefont{Manca, Glazov, Robert,
  Cadiz, Taniguchi, Watanabe, Courtade, Amand, Renucci, Marie
  et~al.}}]{Manca:2017a}
\bibinfo{author}{\bibfnamefont{M.}~\bibnamefont{Manca}},
  \bibinfo{author}{\bibfnamefont{M.~M.} \bibnamefont{Glazov}},
  \bibinfo{author}{\bibfnamefont{C.}~\bibnamefont{Robert}},
  \bibinfo{author}{\bibfnamefont{F.}~\bibnamefont{Cadiz}},
  \bibinfo{author}{\bibfnamefont{T.}~\bibnamefont{Taniguchi}},
  \bibinfo{author}{\bibfnamefont{K.}~\bibnamefont{Watanabe}},
  \bibinfo{author}{\bibfnamefont{E.}~\bibnamefont{Courtade}},
  \bibinfo{author}{\bibfnamefont{T.}~\bibnamefont{Amand}},
  \bibinfo{author}{\bibfnamefont{P.}~\bibnamefont{Renucci}},
  \bibinfo{author}{\bibfnamefont{X.}~\bibnamefont{Marie}},
  \bibnamefont{et~al.}, \bibinfo{journal}{Nature communications}
  \textbf{\bibinfo{volume}{8}}, \bibinfo{pages}{14927} (\bibinfo{year}{2017}).

\bibitem[{\citenamefont{Cadiz et~al.}(2018)\citenamefont{Cadiz, Djeffal,
  Lagarde, Balocchi, Tao, Xu, Liang, Stoffel, Devaux, Jaffres
  et~al.}}]{cadiz2018electrical}
\bibinfo{author}{\bibfnamefont{F.}~\bibnamefont{Cadiz}},
  \bibinfo{author}{\bibfnamefont{A.}~\bibnamefont{Djeffal}},
  \bibinfo{author}{\bibfnamefont{D.}~\bibnamefont{Lagarde}},
  \bibinfo{author}{\bibfnamefont{A.}~\bibnamefont{Balocchi}},
  \bibinfo{author}{\bibfnamefont{B.}~\bibnamefont{Tao}},
  \bibinfo{author}{\bibfnamefont{B.}~\bibnamefont{Xu}},
  \bibinfo{author}{\bibfnamefont{S.}~\bibnamefont{Liang}},
  \bibinfo{author}{\bibfnamefont{M.}~\bibnamefont{Stoffel}},
  \bibinfo{author}{\bibfnamefont{X.}~\bibnamefont{Devaux}},
  \bibinfo{author}{\bibfnamefont{H.}~\bibnamefont{Jaffres}},
  \bibnamefont{et~al.}, \bibinfo{journal}{Nano letters}
  (\bibinfo{year}{2018}).

\bibitem[{\citenamefont{{Kormanyos} et~al.}(2015)\citenamefont{{Kormanyos},
  {Burkard}, {Gmitra}, {Fabian}, {Zolyomi}, {Drummond}, and
  {Fal'ko}}}]{Kormanyos:2015a}
\bibinfo{author}{\bibfnamefont{A.}~\bibnamefont{{Kormanyos}}},
  \bibinfo{author}{\bibfnamefont{G.}~\bibnamefont{{Burkard}}},
  \bibinfo{author}{\bibfnamefont{M.}~\bibnamefont{{Gmitra}}},
  \bibinfo{author}{\bibfnamefont{J.}~\bibnamefont{{Fabian}}},
  \bibinfo{author}{\bibfnamefont{V.}~\bibnamefont{{Zolyomi}}},
  \bibinfo{author}{\bibfnamefont{N.~D.} \bibnamefont{{Drummond}}},
  \bibnamefont{and} \bibinfo{author}{\bibfnamefont{V.}~\bibnamefont{{Fal'ko}}},
  \bibinfo{journal}{2D Materials} \textbf{\bibinfo{volume}{2}},
  \bibinfo{pages}{022001} (\bibinfo{year}{2015}).

\bibitem[{\citenamefont{Cadiz et~al.}(2016{\natexlab{a}})\citenamefont{Cadiz,
  Tricard, Gay, Lagarde, Wang, Robert, Renucci, Urbaszek, and
  Marie}}]{Cadiz:2016a}
\bibinfo{author}{\bibfnamefont{F.}~\bibnamefont{Cadiz}},
  \bibinfo{author}{\bibfnamefont{S.}~\bibnamefont{Tricard}},
  \bibinfo{author}{\bibfnamefont{M.}~\bibnamefont{Gay}},
  \bibinfo{author}{\bibfnamefont{D.}~\bibnamefont{Lagarde}},
  \bibinfo{author}{\bibfnamefont{G.}~\bibnamefont{Wang}},
  \bibinfo{author}{\bibfnamefont{C.}~\bibnamefont{Robert}},
  \bibinfo{author}{\bibfnamefont{P.}~\bibnamefont{Renucci}},
  \bibinfo{author}{\bibfnamefont{B.}~\bibnamefont{Urbaszek}}, \bibnamefont{and}
  \bibinfo{author}{\bibfnamefont{X.}~\bibnamefont{Marie}},
  \bibinfo{journal}{Applied Physics Letters} \textbf{\bibinfo{volume}{108}},
  \bibinfo{pages}{251106} (\bibinfo{year}{2016}{\natexlab{a}}).

\bibitem[{\citenamefont{Cadiz et~al.}(2016{\natexlab{b}})\citenamefont{Cadiz,
  Robert, Wang, Kong, Fan, Blei, Lagarde, Gay, Manca, Taniguchi
  et~al.}}]{cadiz2016ultra}
\bibinfo{author}{\bibfnamefont{F.}~\bibnamefont{Cadiz}},
  \bibinfo{author}{\bibfnamefont{C.}~\bibnamefont{Robert}},
  \bibinfo{author}{\bibfnamefont{G.}~\bibnamefont{Wang}},
  \bibinfo{author}{\bibfnamefont{W.}~\bibnamefont{Kong}},
  \bibinfo{author}{\bibfnamefont{X.}~\bibnamefont{Fan}},
  \bibinfo{author}{\bibfnamefont{M.}~\bibnamefont{Blei}},
  \bibinfo{author}{\bibfnamefont{D.}~\bibnamefont{Lagarde}},
  \bibinfo{author}{\bibfnamefont{M.}~\bibnamefont{Gay}},
  \bibinfo{author}{\bibfnamefont{M.}~\bibnamefont{Manca}},
  \bibinfo{author}{\bibfnamefont{T.}~\bibnamefont{Taniguchi}},
  \bibnamefont{et~al.}, \bibinfo{journal}{2D Materials}
  \textbf{\bibinfo{volume}{3}}, \bibinfo{pages}{045008}
  (\bibinfo{year}{2016}{\natexlab{b}}).

\bibitem[{\citenamefont{Ross et~al.}(2013)\citenamefont{Ross, Wu, Yu, Ghimire,
  Jones, Aivazian, Yan, Mandrus, Xiao, Yao et~al.}}]{Ross:2013a}
\bibinfo{author}{\bibfnamefont{J.~S.} \bibnamefont{Ross}},
  \bibinfo{author}{\bibfnamefont{S.}~\bibnamefont{Wu}},
  \bibinfo{author}{\bibfnamefont{H.}~\bibnamefont{Yu}},
  \bibinfo{author}{\bibfnamefont{N.~J.} \bibnamefont{Ghimire}},
  \bibinfo{author}{\bibfnamefont{A.~M.} \bibnamefont{Jones}},
  \bibinfo{author}{\bibfnamefont{G.}~\bibnamefont{Aivazian}},
  \bibinfo{author}{\bibfnamefont{J.}~\bibnamefont{Yan}},
  \bibinfo{author}{\bibfnamefont{D.~G.} \bibnamefont{Mandrus}},
  \bibinfo{author}{\bibfnamefont{D.}~\bibnamefont{Xiao}},
  \bibinfo{author}{\bibfnamefont{W.}~\bibnamefont{Yao}}, \bibnamefont{et~al.},
  \bibinfo{journal}{Nature communications} \textbf{\bibinfo{volume}{4}},
  \bibinfo{pages}{1474} (\bibinfo{year}{2013}).

\bibitem[{\citenamefont{Dufferwiel et~al.}(2017)\citenamefont{Dufferwiel,
  Lyons, Solnyshkov, Trichet, Withers, Schwarz, Malpuech, Smith, Novoselov,
  Skolnick et~al.}}]{dufferwiel2017valley}
\bibinfo{author}{\bibfnamefont{S.}~\bibnamefont{Dufferwiel}},
  \bibinfo{author}{\bibfnamefont{T.}~\bibnamefont{Lyons}},
  \bibinfo{author}{\bibfnamefont{D.}~\bibnamefont{Solnyshkov}},
  \bibinfo{author}{\bibfnamefont{A.}~\bibnamefont{Trichet}},
  \bibinfo{author}{\bibfnamefont{F.}~\bibnamefont{Withers}},
  \bibinfo{author}{\bibfnamefont{S.}~\bibnamefont{Schwarz}},
  \bibinfo{author}{\bibfnamefont{G.}~\bibnamefont{Malpuech}},
  \bibinfo{author}{\bibfnamefont{J.}~\bibnamefont{Smith}},
  \bibinfo{author}{\bibfnamefont{K.}~\bibnamefont{Novoselov}},
  \bibinfo{author}{\bibfnamefont{M.}~\bibnamefont{Skolnick}},
  \bibnamefont{et~al.}, \bibinfo{journal}{Nature Photonics}
  \textbf{\bibinfo{volume}{11}}, \bibinfo{pages}{497} (\bibinfo{year}{2017}).

\bibitem[{\citenamefont{Back et~al.}(2017)\citenamefont{Back, Sidler, Cotlet,
  Srivastava, Takemura, Kroner, and Imamo{\u{g}}lu}}]{back2017giant}
\bibinfo{author}{\bibfnamefont{P.}~\bibnamefont{Back}},
  \bibinfo{author}{\bibfnamefont{M.}~\bibnamefont{Sidler}},
  \bibinfo{author}{\bibfnamefont{O.}~\bibnamefont{Cotlet}},
  \bibinfo{author}{\bibfnamefont{A.}~\bibnamefont{Srivastava}},
  \bibinfo{author}{\bibfnamefont{N.}~\bibnamefont{Takemura}},
  \bibinfo{author}{\bibfnamefont{M.}~\bibnamefont{Kroner}}, \bibnamefont{and}
  \bibinfo{author}{\bibfnamefont{A.}~\bibnamefont{Imamo{\u{g}}lu}},
  \bibinfo{journal}{Physical review letters} \textbf{\bibinfo{volume}{118}},
  \bibinfo{pages}{237404} (\bibinfo{year}{2017}).

\bibitem[{\citenamefont{Scuri et~al.}(2018)\citenamefont{Scuri, Zhou, High,
  Wild, Shu, De~Greve, Jauregui, Taniguchi, Watanabe, Kim
  et~al.}}]{Scuri:2018a}
\bibinfo{author}{\bibfnamefont{G.}~\bibnamefont{Scuri}},
  \bibinfo{author}{\bibfnamefont{Y.}~\bibnamefont{Zhou}},
  \bibinfo{author}{\bibfnamefont{A.~A.} \bibnamefont{High}},
  \bibinfo{author}{\bibfnamefont{D.~S.} \bibnamefont{Wild}},
  \bibinfo{author}{\bibfnamefont{C.}~\bibnamefont{Shu}},
  \bibinfo{author}{\bibfnamefont{K.}~\bibnamefont{De~Greve}},
  \bibinfo{author}{\bibfnamefont{L.~A.} \bibnamefont{Jauregui}},
  \bibinfo{author}{\bibfnamefont{T.}~\bibnamefont{Taniguchi}},
  \bibinfo{author}{\bibfnamefont{K.}~\bibnamefont{Watanabe}},
  \bibinfo{author}{\bibfnamefont{P.}~\bibnamefont{Kim}}, \bibnamefont{et~al.},
  \bibinfo{journal}{Phys. Rev. Lett.} \textbf{\bibinfo{volume}{120}},
  \bibinfo{pages}{037402} (\bibinfo{year}{2018}).

\bibitem[{\citenamefont{Li et~al.}(2014)\citenamefont{Li, Ludwig, Low,
  Chernikov, Cui, Arefe, Kim, van~der Zande, Rigosi, Hill et~al.}}]{Li:2014a}
\bibinfo{author}{\bibfnamefont{Y.}~\bibnamefont{Li}},
  \bibinfo{author}{\bibfnamefont{J.}~\bibnamefont{Ludwig}},
  \bibinfo{author}{\bibfnamefont{T.}~\bibnamefont{Low}},
  \bibinfo{author}{\bibfnamefont{A.}~\bibnamefont{Chernikov}},
  \bibinfo{author}{\bibfnamefont{X.}~\bibnamefont{Cui}},
  \bibinfo{author}{\bibfnamefont{G.}~\bibnamefont{Arefe}},
  \bibinfo{author}{\bibfnamefont{Y.~D.} \bibnamefont{Kim}},
  \bibinfo{author}{\bibfnamefont{A.~M.} \bibnamefont{van~der Zande}},
  \bibinfo{author}{\bibfnamefont{A.}~\bibnamefont{Rigosi}},
  \bibinfo{author}{\bibfnamefont{H.~M.} \bibnamefont{Hill}},
  \bibnamefont{et~al.}, \bibinfo{journal}{Phys. Rev. Lett.}
  \textbf{\bibinfo{volume}{113}}, \bibinfo{pages}{266804}
  (\bibinfo{year}{2014}).

\bibitem[{\citenamefont{{Lindlau} et~al.}(2017)\citenamefont{{Lindlau},
  {Robert}, {Funk}, {F{\"o}rste}, {F{\"o}rg}, {Colombier}, {Neumann},
  {Courtade}, {Shree}, {Taniguchi} et~al.}}]{2017arXiv171000988L}
\bibinfo{author}{\bibfnamefont{J.}~\bibnamefont{{Lindlau}}},
  \bibinfo{author}{\bibfnamefont{C.}~\bibnamefont{{Robert}}},
  \bibinfo{author}{\bibfnamefont{V.}~\bibnamefont{{Funk}}},
  \bibinfo{author}{\bibfnamefont{J.}~\bibnamefont{{F{\"o}rste}}},
  \bibinfo{author}{\bibfnamefont{M.}~\bibnamefont{{F{\"o}rg}}},
  \bibinfo{author}{\bibfnamefont{L.}~\bibnamefont{{Colombier}}},
  \bibinfo{author}{\bibfnamefont{A.}~\bibnamefont{{Neumann}}},
  \bibinfo{author}{\bibfnamefont{E.}~\bibnamefont{{Courtade}}},
  \bibinfo{author}{\bibfnamefont{S.}~\bibnamefont{{Shree}}},
  \bibinfo{author}{\bibfnamefont{T.}~\bibnamefont{{Taniguchi}}},
  \bibnamefont{et~al.}, \bibinfo{journal}{ArXiv e-prints}
  (\bibinfo{year}{2017}), \eprint{1710.00988}.

\bibitem[{\citenamefont{Zhao et~al.}(2017)\citenamefont{Zhao, Norden, Zhang,
  Zhao, Cheng, Sun, Parry, Taheri, Wang, Yang et~al.}}]{zhao2017enhanced}
\bibinfo{author}{\bibfnamefont{C.}~\bibnamefont{Zhao}},
  \bibinfo{author}{\bibfnamefont{T.}~\bibnamefont{Norden}},
  \bibinfo{author}{\bibfnamefont{P.}~\bibnamefont{Zhang}},
  \bibinfo{author}{\bibfnamefont{P.}~\bibnamefont{Zhao}},
  \bibinfo{author}{\bibfnamefont{Y.}~\bibnamefont{Cheng}},
  \bibinfo{author}{\bibfnamefont{F.}~\bibnamefont{Sun}},
  \bibinfo{author}{\bibfnamefont{J.~P.} \bibnamefont{Parry}},
  \bibinfo{author}{\bibfnamefont{P.}~\bibnamefont{Taheri}},
  \bibinfo{author}{\bibfnamefont{J.}~\bibnamefont{Wang}},
  \bibinfo{author}{\bibfnamefont{Y.}~\bibnamefont{Yang}}, \bibnamefont{et~al.},
  \bibinfo{journal}{Nature nanotechnology} \textbf{\bibinfo{volume}{12}},
  \bibinfo{pages}{757} (\bibinfo{year}{2017}).

\bibitem[{\citenamefont{Zhong et~al.}(2017)\citenamefont{Zhong, Seyler,
  Linpeng, Cheng, Sivadas, Huang, Schmidgall, Taniguchi, Watanabe, McGuire
  et~al.}}]{zhong2017van}
\bibinfo{author}{\bibfnamefont{D.}~\bibnamefont{Zhong}},
  \bibinfo{author}{\bibfnamefont{K.~L.} \bibnamefont{Seyler}},
  \bibinfo{author}{\bibfnamefont{X.}~\bibnamefont{Linpeng}},
  \bibinfo{author}{\bibfnamefont{R.}~\bibnamefont{Cheng}},
  \bibinfo{author}{\bibfnamefont{N.}~\bibnamefont{Sivadas}},
  \bibinfo{author}{\bibfnamefont{B.}~\bibnamefont{Huang}},
  \bibinfo{author}{\bibfnamefont{E.}~\bibnamefont{Schmidgall}},
  \bibinfo{author}{\bibfnamefont{T.}~\bibnamefont{Taniguchi}},
  \bibinfo{author}{\bibfnamefont{K.}~\bibnamefont{Watanabe}},
  \bibinfo{author}{\bibfnamefont{M.~A.} \bibnamefont{McGuire}},
  \bibnamefont{et~al.}, \bibinfo{journal}{Science advances}
  \textbf{\bibinfo{volume}{3}}, \bibinfo{pages}{e1603113}
  (\bibinfo{year}{2017}).

\bibitem[{\citenamefont{Ohuchi et~al.}(1991)\citenamefont{Ohuchi, Shimada,
  Parkinson, Ueno, and Koma}}]{ohuchi1991growth}
\bibinfo{author}{\bibfnamefont{F.}~\bibnamefont{Ohuchi}},
  \bibinfo{author}{\bibfnamefont{T.}~\bibnamefont{Shimada}},
  \bibinfo{author}{\bibfnamefont{B.}~\bibnamefont{Parkinson}},
  \bibinfo{author}{\bibfnamefont{K.}~\bibnamefont{Ueno}}, \bibnamefont{and}
  \bibinfo{author}{\bibfnamefont{A.}~\bibnamefont{Koma}},
  \bibinfo{journal}{Journal of crystal growth} \textbf{\bibinfo{volume}{111}},
  \bibinfo{pages}{1033} (\bibinfo{year}{1991}).

\bibitem[{\citenamefont{Zhang et~al.}(2014)\citenamefont{Zhang, Chang, Zhou,
  Cui, Yan, Liu, Schmitt, Lee, Moore, Chen et~al.}}]{Zhang:2014a}
\bibinfo{author}{\bibfnamefont{Y.}~\bibnamefont{Zhang}},
  \bibinfo{author}{\bibfnamefont{T.-R.} \bibnamefont{Chang}},
  \bibinfo{author}{\bibfnamefont{B.}~\bibnamefont{Zhou}},
  \bibinfo{author}{\bibfnamefont{Y.-T.} \bibnamefont{Cui}},
  \bibinfo{author}{\bibfnamefont{H.}~\bibnamefont{Yan}},
  \bibinfo{author}{\bibfnamefont{Z.}~\bibnamefont{Liu}},
  \bibinfo{author}{\bibfnamefont{F.}~\bibnamefont{Schmitt}},
  \bibinfo{author}{\bibfnamefont{J.}~\bibnamefont{Lee}},
  \bibinfo{author}{\bibfnamefont{R.}~\bibnamefont{Moore}},
  \bibinfo{author}{\bibfnamefont{Y.}~\bibnamefont{Chen}}, \bibnamefont{et~al.},
  \bibinfo{journal}{Nature Nanotechnology} \textbf{\bibinfo{volume}{9}},
  \bibinfo{pages}{111} (\bibinfo{year}{2014}).

\end{thebibliography}
\end{document}